\newcommand{\f}[1]{Fig.~\ref{#1}}

\def\be{\begin{equation}}
\def\ee{\end{equation}}
\def\bea{\begin{eqnarray}}
\def\eea{\end{eqnarray}}
\def\l({\left(}
\def\r){\right)}

\documentstyle[prb,aps,multicol,psfig,graphicx]{revtex}
  \renewcommand{\narrowtext}{\begin{multicols}{2} \global\columnwidth20.5pc}
  \renewcommand{\widetext}{\end{multicols} \global\columnwidth42.5pc}

\begin{document}
{%\large
%%% название
\title{Dendritic flux avalanches in
superconducting Nb$_{3}$Sn films\\[-0.3cm]~\\}

\author{ I.\,A.\,Rudnev
$^a$, S.\,V.\,Antonenko$^a$, D. V. Shantsev$^{b,c}$,
T.\,H.\,Johansen$^{b,}$\cite{1},  A.\,E.\,Primenko$^d$}

\address{$^a$ Moscow Engineering Physics
Institute (State University), 115409 Moscow, Russia\\
$^b$ Department of Physics, University of Oslo, P.O. Box 1048
Blindern, N-0316 Oslo, Norway\\
$^c$ Ioffe Physico-Technical Institute, Polytekhnicheskaya 26,
St.Petersburg 194021, Russia\\
$^d$ Department of Low Temperature Physics and Superconductivity, 
Moscow State University, 117234 Russia}

\maketitle

%\vspace{1cm}
\begin{abstract}
The penetration  of magnetic flux into a thin superconducting film of Nb$_3$Sn
with critical temperature 17.8~K and critical current density 6~MA/cm$^2$
was visualized using magneto-optical imaging.
Below 8~K an avalanche-like flux penetration in form of big and branching 
dendritic structures was observed
in response to increasing perpendicular applied field. 
When a growing dendritic branch meets a linear defect in the film,
several scenarios were observed:
the branch can turn and propagate along the defect,
continue propagation right through it, 
or ``tunnel'' along a flux-filled defect and 
continue growth from its other end. The avalanches manifest themselves in
numerous small and random jumps found in the magnetization curve.  
\end{abstract}

%%% PACS numbers
%\PACS{74.60.-w, 74.60.Ge, 74.70.Ad, 74.76.-w}

%\vspace{1cm}
%\newpage

\narrowtext

The conventional superconductor Nb$_3$Sn belongs to 
the family of A15 materials
characterized by a high value of the critical temperature and upper
critical field, and is therefore
widely used for various technological applications.\cite{book}
One of the central issues is then 
the material's stability with respect to flux jumps, an avalanche
process resulting in a thermal runaway.
Such a dramatic event occurs because flux motion dissipates heat leading
to a local temperature rise, which in turn reduces the flux pinning, 
and thereby
gives a positive feedback to the process.\cite{book,Jc3,Jc4} 

As the critical current density $J_c$ in Nb$_3$Sn 
is constantly being improved,\cite{lee}
the stability criteria for flux jump, e.~g. 
the critical filament width, need to be 
reconsidered.\cite{sumption,collings}
Moreover, there are still open questions related to flux jumping
in general, namely,
that samples often appear more stable experimentally than predicted
by theories.\cite{book}
In investigations of such avalanche events,
one can benefit significantly by  
using thin-film samples, where one can 
visualize the development of the instability by monitoring spatial 
distributions of the magnetic flux.
In this work we used magneto-optical (MO) imaging to study
flux penetration in films of Nb$_3$Sn, which revealed a new type
of instability in this material.

Films of Nb$_3$Sn 
were deposited on 0.5~mm thick sapphire substrates
using magnetron sputtering.\cite{anton}
The films had thickness of 0.1$\div$0.15~$\mu$m. 
The critical temperature and the width of the resistive 
transition measured by a four-probe method were 17.8$\pm$0.1~K.
Critical current densities as high as~6~MA/cm$^2$
were measured at 4.2~K in a magnetic field of 1~T.
For fields below 0.5~T, the critical current density could not be
determined accurately, 
pointing to possible instabilities.\cite{esin}
The presence of flux instabilities becomes evident from the magnetization 
curve, \f{f_m}, showing numerous and random jumps.
The typical jump amplitude here, $\Delta m \sim 0.1 m$, is much larger than the 
sensitivity of the measurements, $3\times 10^{-5}$~emu.   

The spatial distributions of magnetic flux were
visualized using the MO imaging technique\cite{jooss} 
based on 
the Faraday effect in ferrite-garnet indicator films,
see Ref.~\onlinecite{MO1} for
more details on the set-up. The sample, with 
the indicator film placed directly on top, was glued 
to the cold finger of an optical cryostat using vacuum grease. 
The sample was initially cooled to liquid helium temperature in 
zero magnetic field. Then, an applied  
field was turned on perpendicular to the film. 

Shown in \f{f_f1} is
a series of MO images taken in a slowly increasing field. 
Bright and dark regions correspond here 
to high and low flux densities, respectively. 
In (a), which was taken at 5.5~mT, 
one can see that the film is completely shielding the 
applied field. A few exceptions are 
seen near the edge, where due to imperfections
a small amount of flux penetrates in a needle-like
manner. By increasing the field to 8.5~mT
an abrupt event occurs, with 
the flux entering in the form of a branching dendritic structure.
This contrasts the conventional scenario of flux penetration  
where a smooth flux front is gradually advancing in from the 
edges.\cite{jooss}
When the field is increased further, one finds that
many new flux dendrites are abruptly formed, 
and eventually cover most of 
the sample area, as seen in (c) and (d).

\begin{figure}
\centerline{\includegraphics[width=8.2cm]{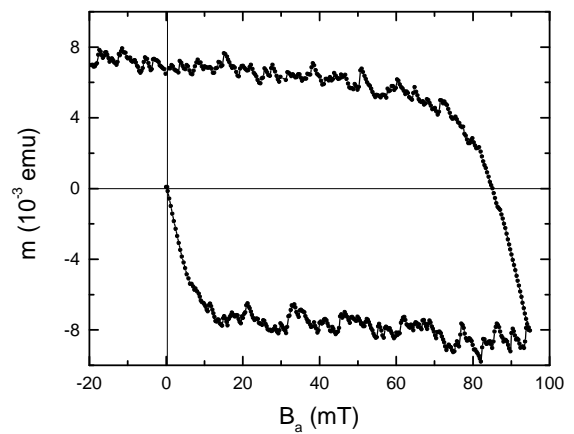}}
\vspace{0.2cm}
\caption{Magnetic moment of a Nb$_3$Sn film at 4.2~K as a 
function of perpendicular applied field. 
The noisy behavior is indicative of local
flux instabilities. The measurements were
taken with a vibrating sample magnetometer PARC, with 
a field ramp rate of 0.15 mT/s.    
\label{f_m}}
\end{figure}

\begin{figure}
\centerline{\includegraphics[width=8.5cm]{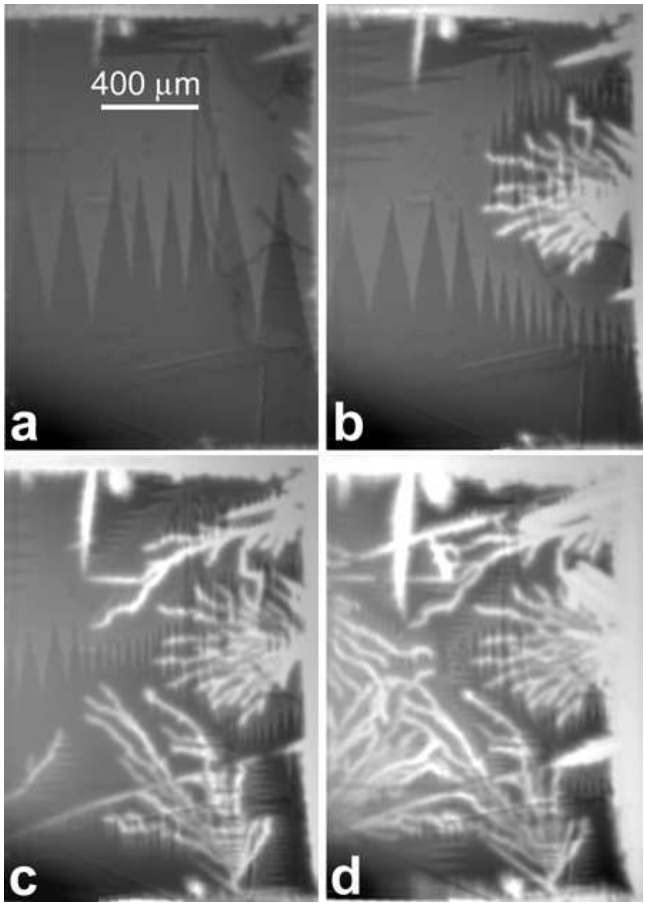}}
%\centerline{\psfig{figure=ha.eps,width=8.5cm}}
\vspace{0.4cm}
\caption{MO images of the flux 
distribution in a Nb$_3$Sn film at $3.5$~K
for increasing magnetic field of 5.5 (a), 8.5 (b), 14.5 (c), and 26.3~mT (d). 
The image brightness represents the magnitude of the
local flux density. The flux penetrates abruptly in the form of 
branching dendritic patterns. 
Zigzag lines are artifacts 
caused by domains in the MO indicator. 
\label{f_f1}}
\end{figure}

\begin{figure}
\centerline{\includegraphics[width=8.5cm]{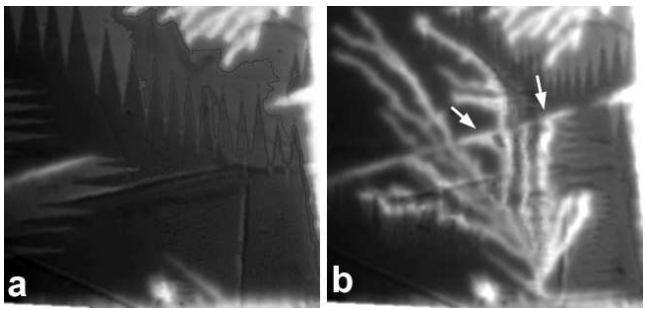}}
%\centerline{\psfig{figure=f2.eps,width=8.5cm}}
\vspace{0.3cm}
\caption{MO images showing penetration of a flux dendrite 
into a region with a linear defect.
The dendrite was abruptly formed as the field increased from 
14~mT (a) to 14.5~mT (b).  Some of dendritic branches (marked by arrows)
are stopped
by the defect, while others penetrate right through.
\label{f_f2}}
\end{figure}

The observed behavior is similar to the dendritic instability
found earlier in films of a few other superconducting materials:
Nb,\cite{Nb1,Nb2,vv} YBa$_2$Cu$_3$O$_{7-x}$ 
(when triggered by a laser pulse),\cite{Y1,Y2}  
and the recently discovered MgB$_2$.\cite{epl,apl,alfoil} 
The common features are that  
the dendrites are formed at seemingly random places along the edge, 
from where they develop extremely fast
(less than 1~ms), and remain afterwards ``frozen'', i.e., 
the grown dendritic structure is not affected by the
subsequent field increase. Furthermore, when new 
dendrites propagate, they tend to avoid  
the already existing ones, as seen in \f{f_f1}(c). 
If the field is later decreased,
new dendrites are formed showing that also the exit of flux from the 
film takes place via dendritic avalanches.
When the experiment is repeated 
at higher temperatures, the number of the dendritic 
structures goes down, and we find  in Nb$_3$Sn that
they disappear completely above 8~K.

It is clear now that the dendritic flux avalanches 
are responsible for the small random jumps in 
magnetization seen in \f{f_m}. 
Remarkably, the field when the first dendrite is formed, 8.5~mT, is in a
very good agreement with the field of the first jump on the virgin branch of $m(H)$.  
The critical current density determined from such $m(H)$ curve 
appears significantly reduced compared to its ``true'' value that it would take
in the absence of instabilities.
The reduction can be as high as 50\% as demonstrated for 
MgB$_2$ films,\cite{epl} where similar noisy magnetization curves
have been measured in a wide range of $T$ and $H$.\cite{zhao}
The dendritic avalanches can also be induced by pulses of 
transport current,\cite{apl} which threatens high-current applications.  
The thermo-magnetic origin of the dendritic instability is 
supported both by
experiments\cite{alfoil} and computer simulations.\cite{aranson,epl}

A specific feature of our Nb$_3$Sn sample is the presence
of linear defects which serve as channels 
of easy flux penetration, see \f{f_f1}. 
These straight bright lines in the MO images
represent regions of suppressed superconductivity, and are probably 
related to defects in the substrate created by imperfect polishing.
Normally, we find that the flux penetration into these defects 
proceeds gradually.
For example, the vertical-line
defect appearing near the middle of the top edge clearly grows
in size as the field increases, see \f{f_f1}(a)-(d).
However, this gradual penetration is sometimes disturbed
by an abrupt invasion of a dendrite. Conversely, also the propagation 
of the dendrite then becomes perturbed
by the defect, and this interesting interplay 
is illustrated in Figs.~\ref{f_f2} and \ref{f_tun}.

The MO image in \f{f_f2}(a) was taken just before a dendritic avalanche
entered the lower part of the film.
Clearly seen is  
a linear defect, which is here only 
partly penetrated by magnetic flux from the right and the left edges.
The image (b) shows the same region right after the dendrite had entered.
We see here two types of behavior when the dendritic 
branches reach the defect.  
Some branches (marked by arrows) suddenly terminate 
their propagation at the defect, and pump it with flux.
Hence, we find here the same tendency as for  
conventional penetration, namely,
that flux prefers to propagate along defects. 
On the other hand, some branches of the dendritic structure actually 
cross the defect as if it was not there.
Based on this, we suggest that the different branches 
of the same dendritic tree do not grow simultaneously.
First, those of the terminating type reach the defect
and fill it with flux. Eventually, this makes the defect
unattractive for the later arriving branches which simply run it over.

Another kind of dendrite--defect interaction is illustrated in \f{f_tun}, 
where
image (a) shows the flux density distribution at 20~mT.
By increasing the field by 0.7~mT, a large dendritic structure penetrated
from the left. The change in the flux distribution is presented
in (b), which was obtained by subtracting two subsequent MO images.
Surprisingly, the bright regions, which show
where the flux arrived, seem disconnected.
This raises the question of how the flux did enter the 
isolated area seen in the upper half of image (b).
By comparing (a) and (b) it appears that the isolated part
is connected to the main dendritic structure through a 
linear defect as indicated by arrows.
We believe that this defect served to mediate
the propagation of this long dendritic branch.
All the flux contained in the isolated part 
has evidently passed through the defect. 
At the same time, the flux density in the defect itself
was not affected by this process since 
it appears black in the difference image.

In conclusion,
macroscopic dendritic flux avalanches were found in Nb$_3$Sn films
subjected to a varying applied field.
These avalanches destroy the critical state and can be a threat for
applications using this material. 
We also demonstrate that 
when a growing flux dendrite meets a flux-free linear defect, it tends to 
propagate along the defect. If the defect is already filled with flux, 
the dendrite can cross the defect or ``tunnel'' through it and 
continue growth at the other end.

The work was supported by the Norwegian Research Council, 
NorFa, and the Russian Federal Program 
``Integration'', project B-0048.    

\begin{figure}
\centerline{\includegraphics[width=8.5cm]{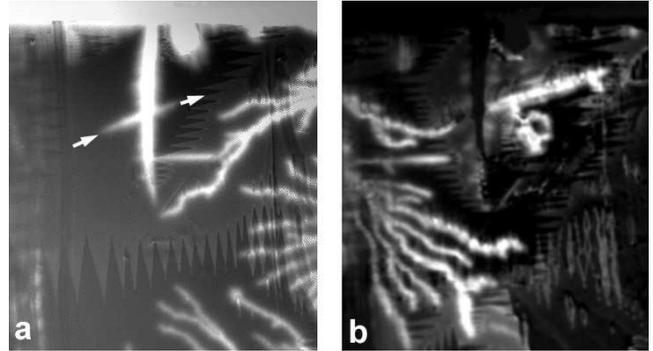}}
%\centerline{\psfig{figure=ha.eps,width=8.5cm}}
\vspace{0.3cm}
\caption{Dendrite propagation mediated by a 
linear defect: (a) MO image taken at an applied field of 20~mT, (b) 
the difference between images taken at 20.7~and 20~mT.
In (b), where the newly formed dendrite appears bright, one of its
branches is disconnected. The flux penetrated into the isolated branch
through the linear defect, as shown by arrows in (a).
\label{f_tun}}
\end{figure}

%\newpage
\widetext

%\end{thebibliography}
\newpage

%\vfill\eject

}

\begin{references}
%\begin{thebibliography}{11}

\vspace{-1.2cm}

\bibitem[*]{1} Email: tomhj@fys.uio.no
\bibitem{book} M. S. Wilson, {\em Superconducting Magnets}. 
Clarendon Press Oxford, 1983.

%\bibitem{Jc1} S.\,L. Wipf, Phys. Rev.  {\bf 161}, 404 (1967).

%\bibitem{Jc2} P.\,S. Swartz, and C.\,P. Bean, 
%J. Appl. Phys. {\bf 39}, 4991 (1968).

\bibitem{Jc3}
S.\,L. Wipf, Cryogenics {\bf 31}, 936 (1991).

\bibitem{Jc4}
R.\,G. Mints, A.\,L. Rakhmanov, Rev. Mod. Phys. {\bf 53}, 551 (1981).

%\bibitem{Jc5}
%A. Vl. Gurevich and R. G. Mints, Rev. Mod. Phys. {\bf 59}, 941 (1987).

\bibitem{lee} P. J. Lee, A. A. Squitieri and D. C. Larbalestier, 
IEEE Trans. Appl. Supercond. {\bf 10}, 979 (2000).

\bibitem{sumption} M.D. Sumption, E. W. Collings, E. Gregory,
IEEE Trans. Appl. Supercond. {\bf 9}, 1455 (1999).

\bibitem{collings} E. W. Collings, M. D. Sumption, E. Lee,
IEEE Trans. Appl. Supercond. {\bf 11}, 2567 (2001).

\bibitem{anton}  S. V. Antonenko and G. A. Komandin,
Prib. Tekh. Eksp. {\bf 4}, 210 (1991), in Russian.

\bibitem{esin} I. A. Esin and I. A. Rudnev, 
%The radiation defects influence on critical 
%current of Nb3Sn superconducting films.  
Fizika Metallov i Metallovedenie,
{\bf 66}, 486 (1988), in Russian.

\bibitem{jooss}
Ch. Jooss, J. Albrecht, H. Kuhn, S. Leonhardt and H. Kronmueller,
Rep. Prog. Phys. {\bf 65}, 651 (2002).

\bibitem{MO1}
T.\,H. Johansen, M. Baziljevich, H. Bratsberg, et al. , Phys. Rev.
B {\bf 54}, 16264 (1996).

\bibitem{Nb1}
M.\,R. Wertheimer, J.\,de G. Gilchrist, J. Phys. Chem Solids {\bf
28}, 2509 (1967).


\bibitem{Nb2}
C.\,A. Duran, P.\,L. Gammel, R.\,E. Miller, D.\,J. Bishop, Phys.
Rev. B {\bf 52}, 75 (1995).

\bibitem{vv} V. Vlasko-Vlasov, U. Welp, V Metlushko, G. W. Crabtree, 
Physica C  {\bf 341-348}, 1281 (2000).

\bibitem{Y1}
P. Leiderer, J. Boneberg, P. Bruell, V. Bujok, S. Herming- haus,
Phys. Rev. Lett. {\bf 71}, 2646 (1993).

\bibitem{Y2}
U. Bolz, J. Eisenmenger, J. Schiessling, B.-U. Runge, P. Leiderer,
Physica B {\bf 284-288}, 757 (2000).

\bibitem{epl} T.H. Johansen, M. Baziljevich, D.V. Shantsev, P.E. Goa, 
Y.M. Galperin, W.N. Kang, H.J.
Kim, E.M. Choi, M.-S. Kim, S.I. Lee, Europhys. Lett. {\bf 59}, 599 (2002).

%\bibitem{sust}
%T.\,H. Johansen, M.Baziljevich, D.\,V. Shantsev et al., Supercond.
%Sci. Technol.{\bf 14}, 726 (2001).

\bibitem{apl} A.V. Bobyl, D.V. Shantsev, T.H. Johansen, 
W.N. Kang, H.J. Kim, E.M. Choi, S.I. Lee,
Appl. Phys. Lett. {\bf 80},  4588 (2002). 

%\bibitem{prb} F. L. Barkov, D. V. Shantsev, T. H. Johansen, P. E. Goa, 
%W. N. Kang, H. J. Kim, E. M. Choi, S. I. Lee, unpublished, 
%cond-mat/0205361.

\bibitem{alfoil} M. Baziljevich, A. V. Bobyl, D.V. Shantsev, E. Altshuler,
T.H. Johansen and S.I. Lee,
Physica C {\bf 369}, 93 (2002). 

%\bibitem{cooley} L. D. Cooley and P. J. Lee, IEEE 
%Trans. Appl. Supercond. {\bf 11}, 3820 (2001).


\bibitem{zhao} Z. W. Zhao, S. L. Li, Y. M. Ni, H. P. Yang, Z. Y. Liu, H. H. Wen,
W. N. Kang, H. J. Kim, E. M. Choi, and S. I. Lee, 
Phys. Rev. B {\bf 65}, 064512 (2002).

%\bibitem{jin} S. Jin, H. Mavoori, C. Bower, R. B. van Dover, Nature {\bf
%411}, 563  (2001).

%\bibitem{dou} S. X. Dou, X. L. Wang, J. Horvat, D. Milliken, A. H. Li, 
%K. Konstantinov, E. W. Collings, M. D. Sumption and H. K. Liu,  
%Physica C {\bf 361}, 79 (2001).

\bibitem{aranson} I. Aranson, A. Gurevich, V. Vinokur, Phys. Rev. Lett.
{\bf 87}, 067003 (2001).

%I.A.Esin , I.A.Rudnev. The radiation defects influence on critical
%current of Nb3Sn superconducting films.  Fizika Metallov i Metallovedenie,
%66(1988)486 ( in Russian)

\end{references}
\end{document}